\documentclass[a4paper]{jpconf}
\usepackage{graphicx}
\usepackage{amsmath}
\usepackage{amssymb}

\def\up{\uparrow}
\def\down{\downarrow}
\def\xc{_{\rm xc}}
\def\rv{{\bf r}}

\def\beq{\begin{equation}}
\def\eeq{\end{equation}}

\begin{document}
\title{Degeneracy and size consistency in electronic density functional theory}

\author{Paola Gori-Giorgi and Andreas Savin}

\address{Laboratoire de Chimie Th\'eorique, CNRS UMR7616 and
Universit\'e Pierre et Marie Curie, 4 Place Jussieu,
F-75252 Paris, France}

\ead{paola.gori-giorgi@lct.jussieu.fr, andreas.savin@lct.jussieu.fr}

\begin{abstract}
The electronic structure calculations based upon energy density functionals 
are highly successful and widely used both in solid state physics and quantum chemistry. 
Moreover, the Hohenberg-Kohn theorems and the Kohn-Sham method provide 
them with a firm basis.
However, several basic issues are not solved, and hamper the progress to achieve high accuracy. In this paper we focus on the conceptual problem of size consistency, basing our analysis on the non-intensive character of the (spin) electronic density in the presence of degeneracy. We also briefly discuss some of the issues concerning fractional electron numbers from the same point of view, analyzing the behavior of the exact functionals for the He and the Hooke's atom 
series when the number of electrons fluctuates between one and two.  
\end{abstract}

\section{Introduction}
Density functional theory (see, e.g., \cite{Koh-RMP-99}) (DFT)  is by now the
most popular method for electronic structure calculations in condensed matter
physics and quantum chemistry, because of its unique combination of low
computational cost and reasonable accuracy for many molecules and solids.
However, despite its large success in scientific areas ranging from material 
science to biology, several basic issues in DFT are still unsolved, 
and hamper futher developments towards high accuracy. Besides, some of these 
issues are often ovelooked or simply ignored. We believe that raising them 
is a necessary step towards further improvement of DFT performances. 
To this purpose, in this paper we concentrate on 
a critical issue in this regard, namely the problem of
size consistency in DFT in the presence of degeneracy.

The Hohenberg-Kohn theorems \cite{HohKoh-PR-64} and the Kohn-Sham method 
\cite{KohSha-PR-65}
provide a firm basis for DFT calculations, in which the 
ground state energy and density of a many-electron system is obtained 
by using energy density functionals that can be rigorously defined.
Unfortunately, the exact definition does not give a prescription that can
 be followed
in practice. Approximations are needed for the exchange-correlation 
energy 
$E\xc[n]$ as a functional of the electronic density 
$n(\rv)$, together with the corresponding exchange-correlation part of the
Kohn-Sham potential, i.e., the functional derivative of $E\xc[n]$ with
 respect to $n(\rv)$. It is worth stressing that
none of the available approximate $E\xc[n]$ satisfies two basic
requirements:
\begin{itemize}
\item to provide reasonable error estimates;
\item to allow systematic improvement, meaning that 
one knows how to reduce the errors by means of a well defined procedure. 
\end{itemize}
Moreover, size consistency in DFT, which
is often taken for granted (at least for systems composed of closed-shell
fragments) can still be an issue, as discussed in the following sections. 
This paper is organized as follows.  
Section \ref{sec_nintensive} discusses the general
form of many approximate energy density functionals, focusing on the
size consistency problem in the Hohenberg-Kohn framework. 
A similar analysis within the Kohn-Sham theory is then carried out 
in Sec.~\ref{sec_KS}. Systems with fractional electron number
are discussed from the same point of view in Sec.~\ref{sec_frac}. The last
Sec.~\ref{sec_conclusions} is devoted to concluding remarks.

\section{Size consistency and density functional theory}
\label{sec_nintensive}
The Hohenberg-Kohn theorems \cite{HohKoh-PR-64} state that the energy of a 
many-electron 
system is a variational functional of the density, 
$E[n]$, given by the sum of a universal part, $F[n]$, and a linear
functional determined by the external potential 
$\hat{V}_{ne}=\sum_i v_{ne}(\rv_i)$ acting on the electrons,
\beq
E[n]=F[n]+\int d\rv\, v_{ne}(\rv)\, n(\rv).
\label{eq_Etot}
\eeq
Although the great success
of DFT comes essentially from the introduction of the Kohn-Sham kinetic energy
functional, orbital-free DFT is appealing for its lower computational cost, and
 is nowadays  an active field of research (see, e.g., \cite{LigCar-HMM-05}). 
It is thus worth to
first analyze the size-consistency issue in the pure Hohenberg-Kohn
framework. In 
orbital-free DFT the universal functional $F[n]$ of
Eq.~(\ref{eq_Etot}) is rewritten as
\beq
F[n]=E_{\rm kxc}[n]+E_{\rm H}[n],
\label{eq_Ekxc}
\eeq
where $E_{\rm H}[n]$ is the usual Hartree classical repulsion energy,
$E_{\rm H}[n]=\frac{1}{2}\int d\rv\int d\rv' n(\rv) n(\rv')|\rv-\rv'|^{-1}$, 
and
$E_{\rm kxc}[n]$ is the remaining part of the energy, the 
kinetic and exchange correlation functional,
that needs to be approximated. The simplest approximations to $E_{\rm kxc}[n]$
have the form
\beq
E_{\rm kxc}^{\rm APPR}[n]=\int d\rv f(n(\rv),|\nabla n(\rv)|,...),
\label{eq_apprgeneral}
\eeq
where the function $f$
is chosen to yield accurate properties for selected systems, e.g., the
energy of the uniform electron gas, and/or to satisfy some known exact 
constraints, like scaling relations.

The requirement of size consistency  for a quantum chemistry 
method is that the result for the energy $E(A+B)$ of two non-interacting 
systems $A$ and $B$
(e.g., which are at infinite distance from each other) be equal to the sum 
of their individual energies, $E(A)+E(B)$. Size consistency is evidently 
crucial when computing dissociation energies.

When approximations of the form of Eq.~(\ref{eq_apprgeneral}) are used, 
one is tempted to believe that size-consistency 
is guaranteed if the function $f$ is intensive, i.e.,  
if its value in the domain of space 
pertaining to system $A$, $\Omega_A$, is not changed by the 
presence of the system $B$ very far from $A$. 
A corresponding statement can be made for
the integration $\Omega_B$ over the region of system $B$. As the integral in 
the composite 
system is the sum over the regions of the individual (sub-)systems,
\beq
\int d\rv f(n(\rv),|\nabla n(\rv)|,...)=\int_{\Omega_A} d\rv f(n(\rv),|\nabla n(\rv)|,...)+
\int_{\Omega_B} d\rv f(n(\rv),|\nabla n(\rv)|,...)
\eeq 
size consistency would be guaranteed (for the Hartree and the external 
potential terms the same partition obviously holds for non-interacting
subsystems). 
The basic belief behind this
statement is that the density itself be an intensive quantity, i.e., that
the total density $n_{A+B}(\rv)$ of the composite system $A+B$ be equal to
\beq
n_{A+B}(\rv)=\left\{
\begin{array}{lr}
 n_A(\rv) & \rv \in \Omega_A \\
 n_B(\rv) & \rv \in \Omega_B 
\end{array}
\right.
\label{eq_nintensive}
\eeq
However, this is not true when one of the two systems (or both) has a
degenerate ground state with different densities. In this case, even
an infinitesimal interaction 
with the other system can select one of the states (or a specific ensemble of
some of the degenerate states). 
This change is not infinitesimal, and depends on the nature of the other
system. Simple examples are diatomic molecules in which the individual atoms
have partially filled shells with $\ell>0$ (e.g. B$_2$, C$_2$, ...).
An even simpler example is one of such atoms (with
degenerate non-spherical densities) perturbed by a proton placed at a large
distance.  
Thus, for degenerate systems, Eq.~(\ref{eq_nintensive}) should be replaced by
\beq
n_{A+B}(\rv)=\left\{
\begin{array}{lr}
  \sum_i w_i(B)\, n_{Ai}(\rv) & \rv \in \Omega_A \\
 \sum_i w_i(A)\, n_{Bi}(\rv) & \rv \in \Omega_B 
\end{array}
\right.
\label{eq_nnonintensive}
\eeq
where $n_{Ai}(\rv)$ and $n_{Bi}(\rv)$ are, respectively, the $i^{\rm th}$
degenerate densities of $A$ and $B$, and the notation $w_i(B)$ (and $w_i(A)$)
explicits the dependence of the weights of each ensemble 
($\sum_i w_i=1$ and $0\le w_i\le 1$) 
on the presence of the other system, even if very far away.
Equation (\ref{eq_nnonintensive}) shows the non intensive character of the
density in the presence of degeneracy.

The {\em exact} $F[n]$, of course, would preserve the degeneracy so that
it would give for any linear combination of the degenerate densities, say, 
$n_{Ai}(\rv)$ the same energy for the system $A$, leading to size consistency.
But none of the available approximate functionals is able to preserve the
degeneracy of the physical system, as none is invariant within the set 
of degenerate
densities. Typically, when treating the isolated 
systems with an approximate functional, one obtains a lower energy
with one of the degenerate densities, or for a particular ensemble. 
This means that the approximation is size consistent only for some very 
specific choices of $A$ and $B$, not in general.  

Notice that here we consider the simple case in which different degenerate
wavefunctions yield  different densities. The interesting case of different 
denegerate
wavefunctions corresponding to the same density is deeply analyzed in 
\cite{CapUllVig-PRA-07}.

\section{Kohn-Sham framework}
\label{sec_KS}
In their foundational work, Kohn and Sham \cite{KohSha-PR-65} 
split the functional 
$E_{\rm kxc}[n]$ of Eq.~(\ref{eq_Ekxc}) into
\beq
E_{\rm kxc}[n]=T_s[n]+E\xc[n],
\eeq
where $T_s[n]$ is the kinetic energy of a system of non-interacting fermions
with density $n(\rv)$ \cite{Lie-IJQC-83},
\beq
T_s[n]=\max_v \left\{\min_\Phi \langle \Phi|\hat{T}+\hat{V}|\Phi\rangle-
\int d \rv\,n(\rv)\,v(\rv)\right\}, \quad {\rm with}\;\hat{V}=\sum_{i=1}^N v(\rv_i),
\label{eq_lieb}
\eeq
where $\Phi$ is in most cases a single Slater determinant. 
The $N$ spin-orbitals $\phi_i$
entering in $\Phi$ are determined via the self-consistent equations
\begin{eqnarray}
\left[-\frac{1}{2}\nabla^2+v_{\rm H}(\rv)[n]+v_{\rm xc}(\rv)[n]+v_{ne}(\rv)\right]
\phi_i(\rv) =  \epsilon_i\phi_i(\rv), & & 
n(\rv)=\sum_i f_i |\phi_i(\rv)|^2, \nonumber \\
v_{\rm H}(\rv)[n]=\frac{\delta E_{\rm H}[n]}{\delta n(\rv)}, \qquad \qquad 
v\xc(\rv)[n]=\frac{\delta E\xc[n]}{\delta n(\rv)}. & & 
\end{eqnarray}
The Kohn-Sham potential $v_{\rm KS}=v_{\rm H}[n]+v_{\rm xc}[n]+v_{ne}$
is the maximizing potential of Eq.~(\ref{eq_lieb}).
Common approximations for the exchange-correlation energy 
$E\xc[n]$ are typically of the form of 
Eq.~(\ref{eq_apprgeneral}), so that as far as size consistency in
the presence of degeneracy is concerned, we can apply 
to the Kohn-Sham $E\xc[n]$ the same considerations 
of the previous section, although in this
case the situation is complicated by the presence of the functional $T_s[n]$, 
which depends on the density in a rather complex way. 
$T_s[n]$ can have different 
values for different densities $n_i(\rv)$ that are
degenerate in the physical system. The work
of Fertig and Kohn \cite{FerKoh-PRA-00} clearly shows that to different
degenerate $n_i(\rv)$ can correspond different Kohn-Sham
potentials $v_{{\rm KS},i}(\rv)$. For example, in an open shell atom with
 several degenerate non-spherical densities we have different non-spherical
Kohn-Sham potentials, one for each symmetry. And for an ensemble of these
densities, we usually need yet another Kohn-Sham potential 
\cite{NagLiuBar-JCP-05}. 

The requirement for approximate $E\xc[n]$ to recover the degeneracy
of the physical system has been recognized by several authors. Usually,
it is written in the form (see, e.g., Ref. 
\cite{YanZhaAye-PRL-00})
\beq
T_s[\Sigma_i w_i\, n_i]+E\xc[\Sigma_i w_i\, n_i]+E_{\rm H}[\Sigma_i w_i\, n_i]=
\Sigma_i w_i \left(T_s[n_i]+E\xc[n_i]+E_{\rm H}[n_i]\right),  
\label{eq_degeneracy}
\eeq
where $n_i$ are the densities of the physical system corresponding
to a set of orthonormalized degenerate ground-state wavefunctions $\Psi_i$, and
Eq.~(\ref{eq_degeneracy}) should hold for any set of the weights $w_i$. 
However,
as noted in Refs.~\cite{Sav-INC-96,NagLiuBar-JCP-05}, imposing
Eq.~(\ref{eq_degeneracy}) to approximate $E\xc[n]$ is probably a 
daunting task: the Hartree energy $E_{\rm H}[\Sigma_i w_i\, n_i]$ contains
cross terms $ij$, and must be compensated by a complex interplay
between $T_s$ and $E\xc$. This is illustrated with simple examples in the
next Sec.~\ref{sec_frac}.

In practical Kohn-Sham calculations, the individual spin densities,
$n_\up(\rv)$ and $n_\down(\rv)$, are used as two indipendent
variables. In this case the total energy is rewritten as
\beq
E[n_\up,n_\down]=T_s[n_\up,n_\down]+E_{\rm H}[n]+E\xc[n_\up,n_\down]+\int d\rv \,v(\rv)\, n(\rv).
\eeq
As the total density, the spin densities are also non-intensive. 
Besides, degeneracy can occur more often than when considering the total density only. The simplest example is the hydrogen atom, which has two degenerate set
of spin densities, $n_\up =n,\; n_\down=0$ and $n_\up =0,\; n_\down=n$. In the
streched hydrogen molecule, we have the equi-ensemble of the two on each atom. 

\section{Fractional number of electrons}
\label{sec_frac}
The behavior of density functionals in the presence of a
fractional number of electrons has been widely investigated
\cite{PerParLevBal-PRL-82,PerLev-PRL-83,ShaSch-PRL-83}. Nowadays,
there is a renovated interest in this issue 
(see, e.g., \cite{MorCohYan-JCP-06,VydScuPer-JCP-07,Peretal-PRA-07}),
which has lead to the definition of the
 ``many-electron self-interaction error''.
\begin{figure}
\includegraphics[width=\columnwidth]{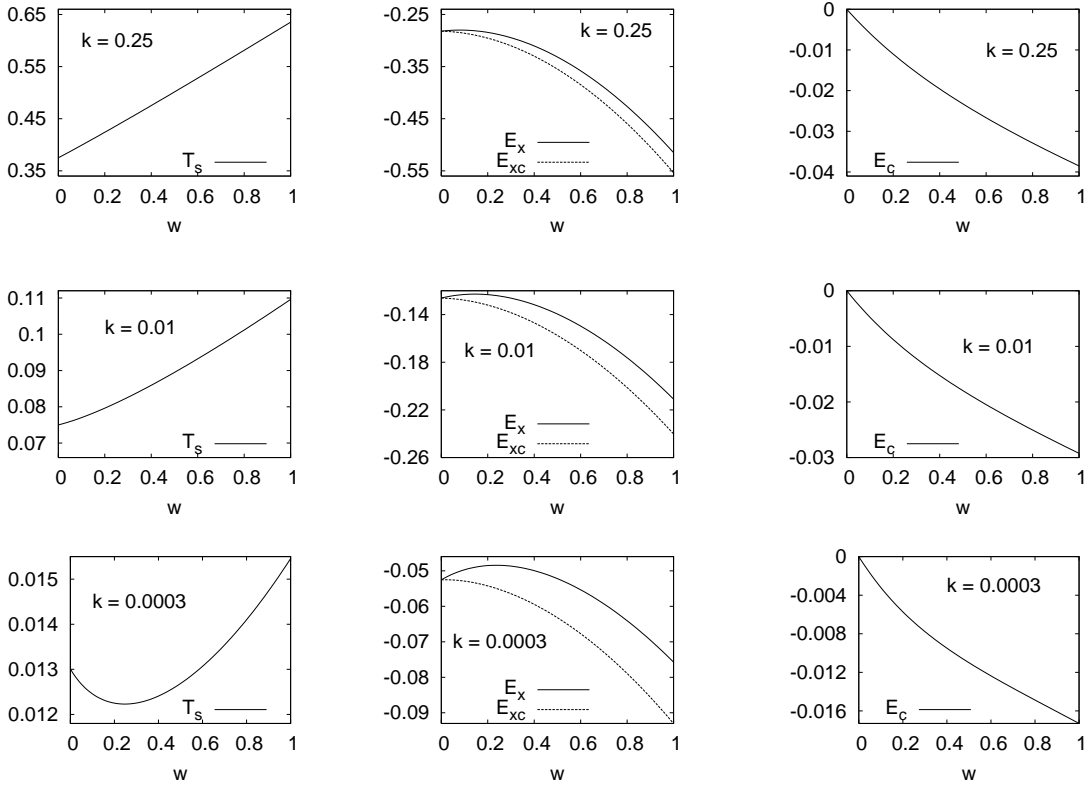}
\caption{\label{fig_hooke}The non-interacting Kohn-Sham kinetic energy
functional $T_s$, the exchange functional $E_{\rm x}$, the 
correlation functional $E_c$, and their sum $E\xc$ for the ensemble
$n=(1-w)\,n_1+w\,n_2$
formed by the $N=1$ and $N=2$ densities of the Hooke's atom series, as
a function of the weight $w$. 
All quantities are exact,
calculated from the analytical solutions given by Taut \cite{Tau-PRA-93}, and are
reported in Hartree atomic units.}
\end{figure}

Fractional number of electrons can be recast in the same class of
problems (degeneracy and size-consistency) discussed in the previous
sections (see also Ref. \cite{YanZhaAye-PRL-00}). 
To understand why, consider the simple example of the stretched
He$_2^+$ molecule, where now system $A$ is the He nucleus plus its electronic
cloud on the left, and system $B$ is the one on the right. We can take the
two degenerate densities $n_{2,1}(\rv)$ and $n_{1,2}(\rv)$,
\beq
n_{2,1}(\rv)=\left\{
\begin{array}{lr}
  n_{\rm He}(\rv) & \rv \in \Omega_A \\
 n_{\rm He^+}(\rv) & \rv \in \Omega_B 
\end{array}
\right.
\qquad
n_{1,2}(\rv)=\left\{
\begin{array}{lr}
  n_{\rm He^+}(\rv) & \rv \in \Omega_A \\
 n_{\rm He}(\rv) & \rv \in \Omega_B 
\end{array}
\right.
\eeq
and form any ensemble of the two. We should always get the same
energy. In particular, we can choose the symmetric one, 
$n_{3/2,3/2}(\rv)=\frac{1}{2}n_{2,1}(\rv)+\frac{1}{2}n_{1,2}(\rv)$, 
which yields $\frac{3}{2}$ 
electrons on $A$ and $\frac{3}{2}$ electrons on $B$. Since the
two degenerate wavefunctions corresponding to $n_{1,2}(\rv)$
and $n_{2,1}(\rv)$ have zero overlap, 
we must have, for any $0\le w\le 1$,
\beq
E[w\, n_{2,1}+(1-w) \,n_{1,2}]=w \,E [n_{2,1}]+ (1-w)\, E[n_{1,2}].
\eeq
If we now only look at the region pertaining to system $A$, we
find \cite{PerParLevBal-PRL-82,YanZhaAye-PRL-00}
\beq
E_A[(1-w) \,n_{\rm He^+}+w\,n_{\rm He}]=(1-w)\, E_A[n_{\rm He^+}]+
w\,E_A[n_{\rm He}],
\eeq
which is the usual result for the energy of a fractional number
of electrons \cite{PerParLevBal-PRL-82,PerLev-PRL-83,ShaSch-PRL-83}, 
corresponding to the well-known requirement for density functionals as
a function of $w$, 
\begin{eqnarray}
T_s[(1-w)\,n_{N}+w\,n_{N+1}]+E_{\rm H}[(1-w)\,n_{N}+w\,n_{N+1}]
+E\xc[(1-w)\,n_{N}+w\,n_{N+1}] = & & \nonumber \\
(1-w)\,\left(T_s[n_N]+E_{\rm H}[n_N]+E\xc[n_N]\right)+w\,
\left(T_s[n_{N+1}]+E_{\rm H}[n_{N+1}]+E\xc[n_{N+1}]\right), & & 
\label{eq_linear}
\end{eqnarray}
where $n_N$ and $n_{N+1}$ are the densities of the physical system (same
$v_{ne}$) with $N$ and $N+1$ electrons, respectively.
Equation~(\ref{eq_linear}) is formally equivalent to Eq.~(\ref{eq_degeneracy}).
We have only two densities $n_N$ and $n_{N+1}$, non-degenerate
on the same region $A$, coming from the degeneracy arising in
systems composed of many sub-systems \cite{YanZhaAye-PRL-00}.

Again, we may wonder whether Eq.~(\ref{eq_linear}) can ever be attained by
approximate functionals. To illustrate the complicated interplay between
$T_s$ and $E\xc$ as a function of $w$, we consider here the simple cases of
the Hooke's
atom series (two interacting electrons in an harmonic external potential,
$v_{ne}(\rv)=\frac{1}{2}k r^2$), and of the He isoelectronic series 
(two interacting electrons with $v_{ne}(\rv)=-Z/r$). 
The Hooke's atom series has a set of analytical solutions 
for specific values of the spring constant $k$ 
\cite{Tau-PRA-93}. Thus, we can calculate the exact densities and energies 
for $N=1$ and $N=2$
electrons, and, by inversion (see \cite{SagPer-PRA-08}), all the exact 
Kohn-Sham functionals
corresponding to the ensemble density $(1-w)\,n_{1}+w\,n_{2}$.
In Figure~\ref{fig_hooke} we report, as a function of $w$, the 
exact $T_s[(1-w)\,n_{1}+w\,n_{2}]$ and $E\xc[(1-w)\,n_{1}+w\,n_{2}]$ (with
its two components, exchange and correlation, separately) for three
different values of $k$. As $k$ decreases, the system becomes more and more correlated. The non-interacting kinetic energy $T_s$ as a function of $w$ changes from being almost linear in the less correlated case $k=\frac{1}{4}$, to displaying a minimum in the more correlated case $k=4\left(\frac{35-3\sqrt{57}}{1424}\right)^2\approx 0.0003$. The functional $E\xc$ must compensate the quadratic
behavior of $E_{\rm H}$ with $w$, 
\beq
E_{\rm H}[(1-w)\,n_{1}+w\,n_{2}]=(1-w)^2\,E_{\rm H}[n_1]+w^2\,E_{\rm H}[n_2]+
w\,(1-w)\,\int d\rv\int d\rv' \frac{n_1(\rv)n_2(\rv')}{|\rv-\rv'|},
\label{eq_EhQuad} 
\eeq
as well as the non-linear behavior of $T_s$ for correlated systems. 

\begin{figure}
\includegraphics[width=\columnwidth]{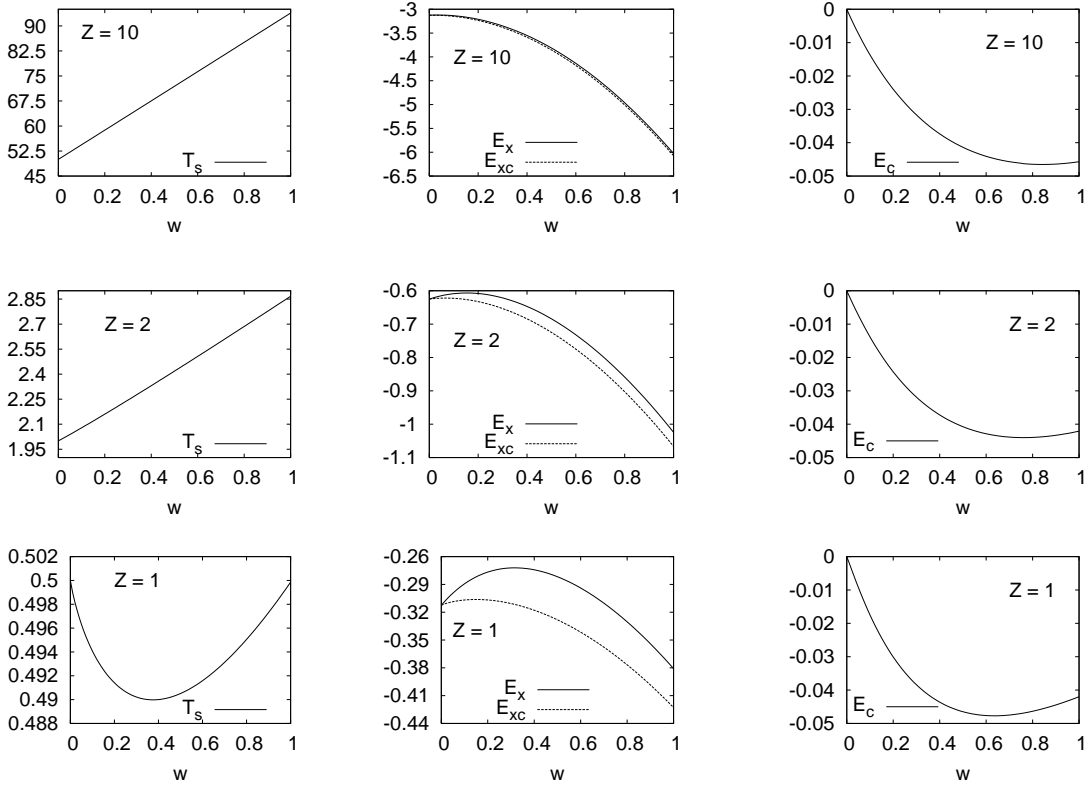}
\caption{\label{fig_he}The non-interacting Kohn-Sham kinetic energy
functional $T_s$, the exchange functional $E_{\rm x}$, the 
correlation functional $E_c$, and their sum $E\xc$ for the ensemble
$n=(1-w)\,n_1+w\,n_2$
formed by the $N=1$ and $N=2$ densities of the He atom series, as
a function of the weight $w$. 
All quantities are very accurate,
calculated from the variational wavefunctions 
of Ref.~\cite{FreHuxMor-PRA-84} (see
also \cite{UmrGon-PRA-94} and \cite{GorSav-PRA-05}), and are
reported in Hartree atomic units.}
\end{figure}

A very similar trend of the functionals is observed for the He
isoelectronic series, as shown in Fig.~\ref{fig_he}. In this case we
have used an improved version~\cite{UmrGon-PRA-94} of the accurate 
variational wavefunctions and energies of
 Ref.~\cite{FreHuxMor-PRA-84} (see also
\cite{GorSav-PRA-05}) to calculate the exact functionals. Notice in particular
the behavior of $T_s$ for the more correlated case, $Z=1$.

A closed shell interacting system of two-electrons is weakly correlated when 
$n_2(\rv) \approx 2 n_1(\rv)$. In this case, the ensemble density with
the corresponding $N=1$ system 
is simply given by 
$w\,n_1+(1-w)\,n_2\approx (1+w)\,n_1$, and $T_s[(1-w)\,n_1+w\,n_2]\approx
(1-w)\,T_s[n_1]+w \,T_s[n_2]$. This is the case, e.g., of the Ne$^{8+}$
system of Fig.~\ref{fig_he}, for which $T_s$ is almost linear.
However, the deviation of the Hartree term from linearity,
\begin{eqnarray}
&  E_{\rm H}[(1-w) n_1+w n_2]-(1-w) E_{\rm H}[n_1]-w E_{\rm H}[n_2] \nonumber = \\
 = &  -
\frac{w\,(1-w)}{2}\int d\rv \int d\rv'\frac{[n_2(\rv)-n_1(\rv)][n_2(\rv')-n_1(\rv')]}{|\rv-\rv'|}=-w\,(1-w)\,E_{\rm H}[n_2-n_1]
\end{eqnarray}
is significantly different from zero
when $n_2\approx 2 n_1$, while
it could become smaller when
$n_2$ is much more diffuse than $2\, n_1$. 
For example, for an ensemble of H and H$^{-}$ we have $E_{\rm H}[n_2-n_1]=
0.1203$ Hartree, while $E_{\rm H}[n_1]=0.3125$ Hartree. That is, if
the $N=2$ system were less correlated so that $n_2-n_1$ were equal (or close) 
to $n_1$,
$E_{\rm H}$ would be further from the linear behavior. So (in the
simple systems considered here) when  $T_s$ is
closer to linearity $E_{\rm H}$ can be further from it, and viceversa. 

While the relative deviation of $T_s$ from linearity decreases as the 
$N=2$ system becomes less correlated (as shown in Figs.~\ref{fig_hooke}
and \ref{fig_he}), the maximum absolute value of 
$T_s[(1-w) n_1+w n_2]-(1-w) T_s[n_1]-w T_s[n_2]$
is always of the same
 order of magnitude, i.e. $\approx 2$ mH for the Hooke's series and 
$\approx 15$ mHartree for the He series.


\section{Conclusions}
\label{sec_conclusions}
Size-consistency in DFT is often taken for granted with approximate functionals
of the form of Eq.~(\ref{eq_apprgeneral}), because the density is believed to
be an intensive quantity. However, the density is not intensive in the presence of degeneracy. An attempt to build a correct description of ensembles in DFT seems in order \cite{UllKoh-PRL-01}, but using 
Eqs.~(\ref{eq_degeneracy}) and (\ref{eq_linear}) is probably not the best
starting point. The alternative path of building ensembles of Kohn-Sham
systems (e.g., using different Kohn-Sham potentials for each symmetry 
of the degenerate system \cite{Sav-INC-96})  deserves further investigation.

\ack
It is our pleasure to dedicate this paper to Cesare Pisani 
with admiration for his constant strive to deep understanding.


\providecommand{\newblock}{}

\end{document}